\documentclass[showpacs,amsmath,amssymb,aps,twocolumn,longbibliography,pra]{revtex4-2}
\usepackage{graphicx}
\usepackage[colorlinks=true,citecolor=blue,linkcolor=magenta]{hyperref}
\usepackage[usenames]{color}
\usepackage{relsize}
\usepackage[english]{babel}
\usepackage{enumerate}
\usepackage{bbm}
\usepackage{xcolor}
\usepackage{framed}
\usepackage{bm}
\usepackage{bbm}

\newcommand{\ket}[1]{\left| #1 \right>} 
\newcommand{\bra}[1]{\left< #1 \right|} 

\newcommand {\grsim} {\ {\raise-.5ex\hbox{$\buildrel>\over\sim$}}\ }
\newcommand {\lessim} {\ {\raise-.5ex\hbox{$\buildrel<\over\sim$}}\ }

\usepackage{xcolor}
\usepackage{siunitx}
\usepackage{booktabs}

\newcommand{\nocontentsline}[3]{}
\newcommand{\tocless}[2]{\bgroup\let\addcontentsline=\nocontentsline#1{#2}\egroup}

\newcommand{\RN}[1]{%
  \textup{\uppercase\expandafter{\romannumeral#1}}%
}

\usepackage{xr}
\begin{document}

\title{Bayesian post-correction of non-Markovian errors in bosonic lattice gravimetry}
\author{Bharath Hebbe Madhusudhana$^{1,2}$ }
\author{Andrew K. Harter$^{1}$}
\author{Avadh Saxena$^{3}$ }

\affiliation{$^{1}$MPA-Quantum, Los Alamos National Laboratory, Los Alamos, NM 87545, United States}
\affiliation{$^{2}$New Mexico Consortium, Los Alamos, NM 87544, United States}
\affiliation{$^{3}$Theoretical Division, Los Alamos National Laboratory, Los Alamos, NM 87545, United States}


\begin{abstract} 
We study gravimetry with bosonic trapped atoms in the presence of random spatial inhomogeneity. The errors  resulting from a random, shot-to-shot fluctuating spatial inhomogeneity are quantum non-Markovian. We show that in a system with $L>2$ modes (i.e., trapping sites), these errors can be post-corrected using a Bayesian inference. The post-correction is done via in situ measurements of the errors and refining the data-processing according to the measured error. We define an effective Fisher information $F_{\text{eff}}$ for such measurements with a Bayesian post-correction and show that the Cramer-Rao bound for the final precision is $\frac{1}{\sqrt{F_{\text{eff}}}}$. Exploring the scaling of the effective Fisher information with the number of atoms $N$, we show that it saturates to a constant when there are too many sources of error and too few modes. That is, with $\ell$ independent sources of error, we show that the effective Fisher information scales as $F_{\text{eff}} \sim \frac{N^2}{a+bN^2}$ for constants $a, b>0$ when the number of modes is small: $L<\ell+2$, even after maximization over the Hilbert space. With larger number of modes, $L\geq \ell+2$, we show that the effective Fisher information has a Heisenberg scaling $F_{\text{eff}}=\mathcal O(N^2)$ when optimized over the Hilbert space. Finally, we study the density of the effective Fisher information in the Hilbert space and show that when $L\geq \ell+2$, almost any Haar random state has a Heisenberg scaling, i.e.,  $F_{\text{eff}}=\mathcal O(N^2)$. Based on these results, we develop a Loschmidt echo like experimental sequence for error mitigated gravimetry and gradiometry and discuss potential implementations. Finally, we argue that the effective Fisher information can be interpreted as the Fisher information corresponding to an equivalent non-Hermitian evolution. 
\end{abstract}

\maketitle

\section{Introduction}

With the rapid progress in quantum hardware technologies, achieving a quantum advantage in applications such as quantum sensing is more likely than ever. However, a definitive practical advantage has largely remained elusive, with a few notable exceptions including LIGO~\cite{Aasi:2013}. The major bottleneck hindering the scalability and field deployment of these sensors is their inherent susceptibility to environmental noise and operational errors. As hardware capabilities continue to mature, the development of robust error mitigation and correction protocols specifically tailored for sensing is gaining significant relevance. 
\begin{figure}[h!]
	\includegraphics[width=0.44\textwidth]{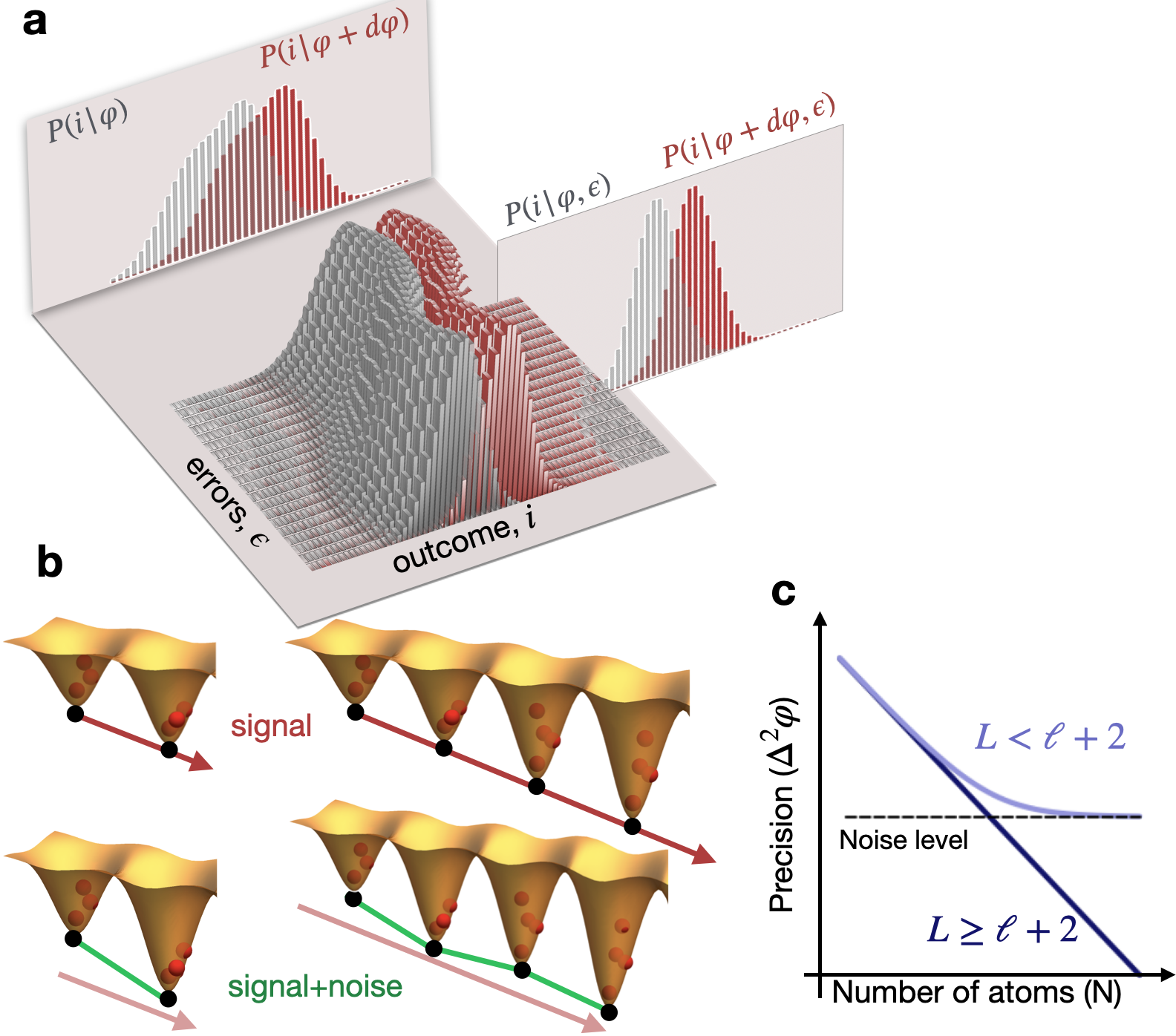}
\caption{\textbf{In situ error detection in quantum sensing:} \textbf{a} Effect of random errors ($\bm{\epsilon}$) on the precision of estimating an unknown parameter $\varphi$.  Each datapoint $i$ is  sampled from a conditional $P(i|\varphi, \bm{\epsilon})$. If the errors are unknown, one has access only to the average distribution $P(i|\varphi)$, which is less sensitive than the conditional to the unknown $\varphi$. \textbf{b.} Errors in a system of two-mode bosons are indistinguishable from the signal; in multi-mode bosons, where we can simultaneously readout the occupancies $\hat{n}_i$, the ``extra information" can be used to detect and post correct the errors. \textbf{c.} Post corrected precision $\Delta^2\varphi$ saturates to a constant, if there are too many sources of error ($\ell$) and too few modes ($L$) to correct them all.  We show that if $L\geq \ell+2$, this precision has a Heisenberg scaling $\sim\frac{1}{N^2}$ with the atom number. }\label{Fig1}
\end{figure}

While quantum error correction applied to computation is a vibrant and growing area, its applications to quantum sensing has received sparse attention so far. One of the central results in this area is the ``Hamiltonian not in Lindblad Space (HNLS)" criterion~\cite{PRXQuantum.2.010343, PhysRevX.7.041009, Sekatski2017quantummetrology}. The latter is a necessary and sufficient criterion for the existence of quantum error correction codes that would recover the Heisenberg limit in the presence of Markovian noise. However, the practical application of this criterion requires noiseless ancilla qubits. 

Other results~\cite{PhysRevLett.122.040502} have shown that error correction protocols that do not use ancilla qubits exist, albeit for a special class of errors. More recently, the metrological performance of more realistic protocols was analyzed~\cite{PhysRevLett.133.170801}, showing that a dephasing error precludes a quantum advantage in almost all of these protocols. On the experimental side, an error correction technique called erasure conversion, which has been used very successfully in quantum computation was applied to quantum sensing recently~\cite{PhysRevLett.133.080801}.

The conceptual tools for inference in quantum metrology remained largely frequentist,  despite the tussle between Bayesian and frequentist inferences in general estimation theory being a few centuries old~\cite{264d98b1-8168-3b24-b814-f5d38293dff8}.  While Bayesian inference in quantum metrology is conceptually different from frequentist inference ~\cite{Li_2018}, the former gained relevance with mid-circuit measurements becoming a possibility.  Since then, Bayesian inference has been used to develop adpative quantum sensing protocols~\cite{StrategyOptimization2026, Belliardo2024, Andre2026} and experimental implementations~\cite{Berni2015, Kanno_2025, youssry2026bayesianquantumsensingusing}.

Most of the work so far has been focused on Markovian errors, while non-Markovian errors are known to also have a significant impact on the scalability of entangled quantum sensors. Moreover, experimental work so far has been scarce and involves smaller number of qubits/atoms shedding no light on the impact of error correction on scalability. Finally, current works use a qubit-based architecture, whereas, many quantum sensing applications use bosons (e.g. Bose-Einstein Condensates, photons, etc).  ~\cite{Gross_2010, PhysRevLett.109.253605, PhysRevLett.113.103004, Hosten_2016, Greve_2022}.

In this work, we show that multi-mode bosonic quantum systems can be engineered so as to detect some of the errors in situ, i.e., simultaneously along with the signal to be measured. We also show that a Bayesian post-correction technique can be applied on the so-detected error recovers the Heisenberg scaling in many experimentally relevant errors. 

We consider sensing an unknown phase $\varphi$, which we refer to as the signal, using $N$ bosons in $L$ modes. The signal can be an external field that is of interest --- we develop the results in this paper assuming it is a gravitational field. We also assume that the errors during the quantum sensing sequence are non-Markovian, i.e., they vary randomly across data points. A typical data point is a sample from a conditional probability distribution $P(i|\varphi, \bm{\epsilon})$, where $i$ represents the possible outcomes, i.e., values of the data point and $\bm{\epsilon}$ represents the error that occurred when \textit{that data point} was taken.  The core idea is that when the errors are unknown and random, the data is really sampled from the \textit{averaged} (over the error) distribution $P(i|\varphi)$ (Fig.~\ref{Fig1}a). When errors are detected in situ, the data is sampled from the \textit{conditional} $P(i|\varphi, \bm{\epsilon})$, which has a lower width in general and is therefore more sensitive to the signal (Fig.~\ref{Fig1}a). 

The key insight for in situ error detection is, in a system of $N$ bosons in $L$ modes, one can \textit{simultaneously} read out the occupancies $\hat{n}_1, \cdots, \hat{n}_L$, which when $L>2$ is more information than just the signal and the ``extra" information can be used to detect some of the errors. Due to the constraint $\hat{n}_1+\cdots+\hat{n}_L=N$, the set of operators $\{\hat{n}_1, \cdots, \hat{n}_L\}$ lie in an $L-1$-dimensional trapezoid, i.e., can hold $L-1$ free parameters. In particular, when $L=2$,  it is one-dimensional and therefore can provide us only one piece of information that includes the signal $\varphi$ and the error $\bm{\epsilon}$ whereas when $L>2$, the larger dimension allows for independent, simultaneous detection of errors along with the signal (see Fig.~\ref{Fig1}b).

We show that the quantum states of a multi-mode bosonic system can be engineered to detect certain forms of error alongside the signal, which can be post-corrected using a Bayesian inference. 

\begin{figure}
    \centering
    \includegraphics[width=0.99\linewidth]{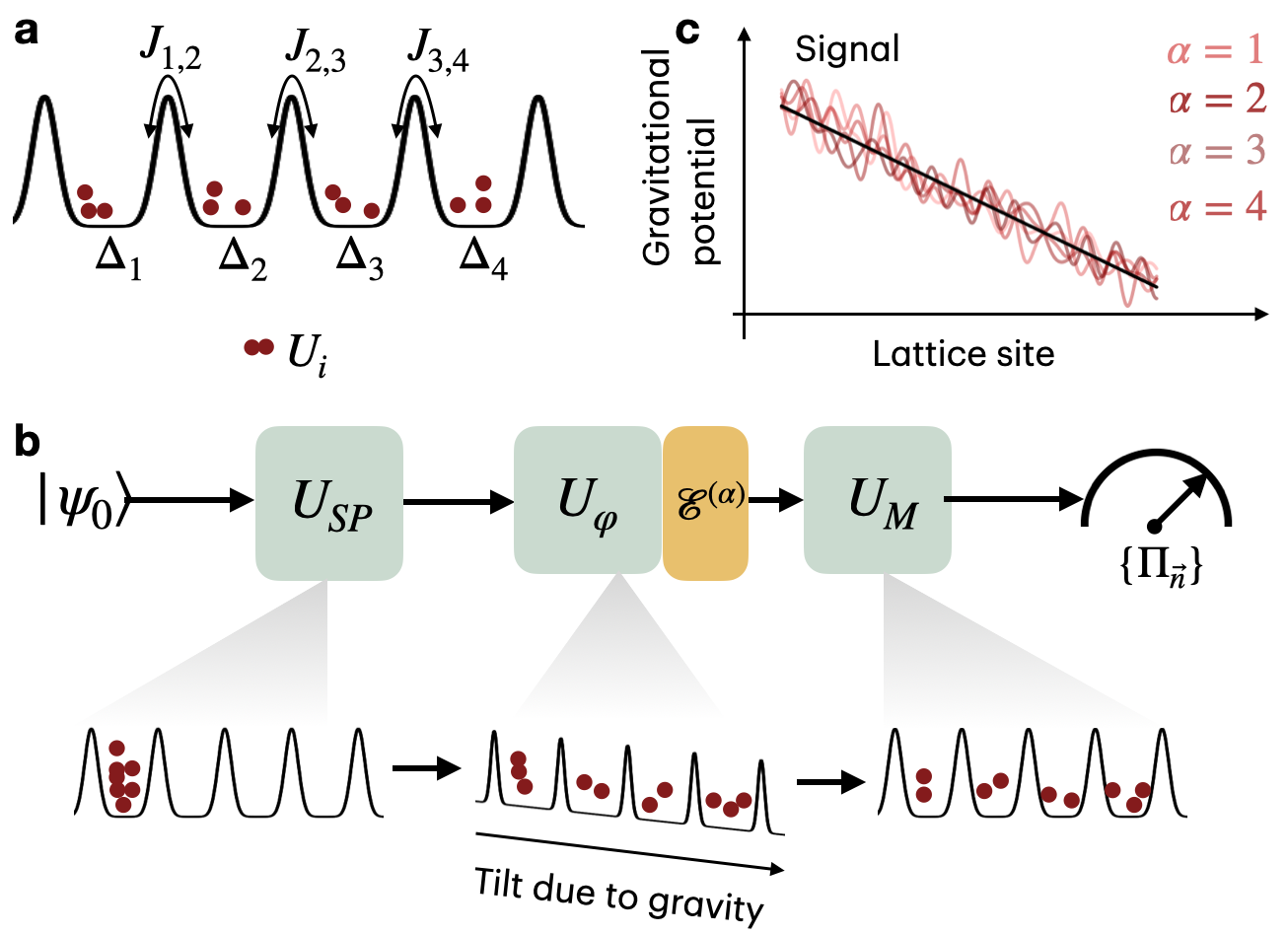}
    \caption{\textbf{System and setup:} \textbf{a.} Controls available in multi-mode bosons, i.e., BEC trapped in an optical lattice (see text). \textbf{b.} Schematic of noisy quantum sensing, with the relevant example of gravimetry with multi-mode bosons.  \textbf{c.} Example of noise in the on-site potential.}
    \label{Fig2}
\end{figure}

\section{Problem setup}

Trapped ultracold atoms have been a very successful platform for quantum sensing~\cite{PhysRevLett.132.190001, Agarwal_2025}.  We consider an atomic BEC with $N$ atoms in an optical lattice with $L$ sites, where each site can be occupied by multiple bosons (Fig.~\ref{Fig2}a,b). Hereafter, $\hat{a}_i^{\dagger}$ for $i=1, \cdots, L$ is the creation operator corresponding to site $i$. We assume that the Hamiltonian of the system is
\begin{equation}\label{Hctrl}
\begin{split}
    H_{\text{ctrl}}(\{J_{i,i+1}, \Delta_i, U_i\}) &= \sum_{i=1}^{L-1} J_{i, i+1} (\hat{a}_i^{\dagger} \hat{a}_{i+1} + \hat{a}_{i+1}^{\dagger} \hat{a}_i)\\
    &+ \sum_{i=1}^L \Delta_ i \hat{a}_i^{\dagger} \hat{a}_i+ \sum_{i=1}^L U_i \hat{n}_i(\hat{n}_i-1) \,. 
\end{split}
\end{equation}
Here, $\hat{n}_i = \hat{a}_i^{\dagger}\hat{a}_i$ is the occupation number of the $i$-th site.  $U_i$ is the site-dependent interaction strength. $J_{i,i+1}$ are the site-dependent hopping rates and $\Delta_i$ are the on-site potentials. We assume that the parameters $J_{i,i+1}, \Delta_i$ and $U_i$ can be controlled independently~\cite{shao2025engineeringmultimodebosonicsqueezed}. (Fig.~\ref{Fig2}b).

A gravitational field along the lattice introduces a ``tilt'', which we intend to measure.  Hamiltonian corresponding to this tilt is
\begin{equation}\label{Htilt}
    H_0 = \eta \sum_{i=1}^L i \hat{a}_i^{\dagger} \hat{a}_i \ .
\end{equation}
Here, $\eta$ is a constant that depends on the strength of the gravitational field. Evolution under this Hamiltonian over time $\tau$ will imprint a  phase $\phi =\eta \tau$ into the quantum state of the system and therefore an estimate of this phase can be used to estimate the parameter $\eta$. 

Let us consider a standard quantum sensing sequence to estimate $\varphi$. This includes $5$ steps (Fig.~\ref{Fig2}a):  (i) the atoms are initialized in a state $\ket{\psi_0}$, typically $\ket{\psi_0}=\frac{1}{\sqrt{N!}}(\hat{a}_1)^N\ket{\text{vac}}$, (ii) a set of quantum control operations, using the Hamiltonian Eq.~(\ref{Hctrl}) are applied on $\ket{\psi_0}$ to prepare a metrologically useful state (e.g., squeezed state), $U_{SP}\ket{\psi_0}$. Here $U_{SP}$ is the state-preparation unitary operator. See ref.~\cite{shao2025engineeringmultimodebosonicsqueezed} for optimization techniques to prepare a squeezed state. (iii) The quantum system interacts with the gravitational field to acquire the phase $\varphi$, described by a unitary $U_{\varphi}=e^{-i\varphi H_0}$. (iv) A set of pre-measurement unitary operations described by another unitary $U_M$ ``prepare" the  measurement basis. (v) Finally, the state is read out, described by a projection valued measure (PVM), $\{\Pi_1,\cdots, \Pi_d \}$. The readout measures the number of atoms $n_i$ in each site, for $i=1,\cdots, L$. Therefore, the PVM is described by $\{\Pi_{\vec{n}} =\ket{\vec{n}}\bra{\vec{n}}:\  \vec{n}=(n_1, \cdots, n_L)\}$. Here, $\ket{\vec{n}}$ is the basis element with $n_i$ atoms in the $i$-th site.

The measurement outcome is a sample from a distribution $P(\vec{n}|\varphi_0) = ||\Pi_{\vec{n}}U_Me^{-i\varphi H_0}U_{SP}\ket{\psi_0}||^2$. Note that $P(\vec{n}|\varphi)$ is the conditional probability distribution. One can use $\nu$ such samples to retrieve an estimate of the parameter $\varphi$, the precision of which is determined by the classical Fisher information (CFI) of $\{P(\vec{n}|\varphi_0)\}$ w.r.t $\varphi$. The unitary $U_M$ is chosen so that the CFI equals the quantum Fisher information (QFI) $F_Q(U_{SP}\ket{\psi_0}, H_0)$. The simplest choice is $U_M=U_{SP}^{-1}$, i.e., a Loschmidt echo. 

We absorb the unitary $U_M$ into the PVM by defining $\tilde{\Pi}_{\vec{n}} = U_M^{\dagger}\Pi_{\vec{n}} U_M$ so that 
\begin{equation}
    P(\vec{n}|\varphi) = ||\tilde{\Pi}_{\vec{n}}e^{-i\varphi H_0}U_{SP}\ket{\psi_0}||^2 \,. 
\end{equation}

Below we describe an error model for quantum sensing.

\subsection{Error model}
Errors can (and do) appear at all $5$ stages described above. Errors in initializing the state $\ket{\psi_0}$ and the final readout are the so-called state preparation and measurement (SPAM) errors. Note that despite the name SPAM, these are \textit{not} related to errors in $U_{SP}$ and $U_M$. We assume that SPAM are negligible in this paper. There can be control errors during $U_{SP}$ and $U_M$, that typically include random events that vary between experimental shots. Here we assume an echo sequence -- $U_M=U_{SP}^{-1}$, so that errors in the application of $U_{SP}$ can be detected by classically monitoring the control parameters in real time and compensated for during the application of $U_M$. For instance, the lasers used to produce $J_{i, i+1}$ can be monitored in real time and errors in it can be compensated for. We focus on errors during the phase acquisition, which we denote by $\mathcal E^{(\alpha)}$, where $\alpha$ is the datapoint number, indicating that the errors vary randomly across datapoints.  The conditional probability is now different for different datapoints and depends on the error 
\begin{equation}
\begin{split}
    P^{(\alpha)}(\vec{n}|\varphi) =
    ||\tilde{\Pi}_{\vec{n}} \mathcal E^{(\alpha)} e^{-i\varphi H_0}U_{SP}\ket{\psi_0}||^2 \,. 
\end{split}
\end{equation}


A few classes of errors have been observed: (1) Markovian errors, e.g., the atoms, they ``leak'' out of the trap or experience internal decay. (2) Non-Markovian errors, e.g., fluctuations in the quantum control between samples and (3) coherent errors, e.g. the presence of additional fields that the system couples to, spurious interactions between atoms, including three-body collisions. In this work, we restrict to coherent and non-Markovian errors. That is, we assume that the error can be described by a Hamiltonian 
\begin{equation}
    H= H_0 + \epsilon_1 H_1 + \epsilon_2 H_2 +\cdots + \epsilon_{\ell}H_{\ell} \,, 
\end{equation}
where $\bm{\epsilon}=(\epsilon_1, \cdots, \epsilon_{\ell})$ are coupling strengths of the errors and $H_1, \cdots, H_{\ell}$ describe the nature of the couplings. For instance, alongside a constant gravitational field $\eta$ if there are other contributions to onsite potentials, modeled by errors $\epsilon_i$ acting on site $i$, $H=H_0 + \sum_i \epsilon_i \hat{n}_i$. This could be other sources of gravity that produce a spatially varying field, etc. The error in $\alpha$-th datapoint is determined in general by a random vector $\bm{\epsilon}^{(\alpha)}=(\epsilon_1^{(\alpha)}, \cdots, \epsilon_{\ell}^{(\alpha)})^T$  sampled from an underlying error distribution $P_{\text{err}}(\bm{\epsilon})$. 
The error would be $\mathcal E^{(\alpha)}e^{-i\varphi H_0} = e^{-i\tau(\eta H_0 + \epsilon_1^{(\alpha)}H_1+\cdots+\epsilon_{\ell}^{(\alpha)}H_{\ell})}$. The probability $P^{(\alpha)}(\vec{n}|\varphi)$ can now be re-written as a conditional $P(\vec{n}|\varphi, \bm{\epsilon}^{(\alpha)} )$:
\begin{equation}
\begin{split}
    P^{(\alpha)}(\vec{n}|\varphi) =&P(\vec{n}|\varphi, \bm{\epsilon}^{(\alpha)} )\\
    = &  ||\tilde{\Pi}_{\vec{n}}  e^{-i(\varphi H_0+\epsilon_1^{(\alpha)}\tau H_1 +\cdots +\epsilon_{\ell}^{(\alpha)}H_{\ell} )}U_{SP}\ket{\psi_0}||^2 \,. 
\end{split}
\end{equation}

A dataset with $\nu$ data points includes outcomes $\{\vec{n}_1, \vec{n}_2, \cdots, \vec{n}_{\nu}\}$ where each $\vec{n}_{\alpha}\in \mathcal B$ is sampled from the corresponding distribution $P(\vec{n}|\varphi, \bm{\epsilon}^{(\alpha)})$. Note that $\bm{\epsilon}^{(\alpha)}$ represents the error that occurred during the $\alpha$-th data point and is a sample from $P_{\text{err}}(\bm{\epsilon})$. It is natural to apply a Bayesian inference to estimate $\varphi$. In the following section, we discuss the error in the Bayesian estimate. 

\section{Bayesian inference and errors}

The Bayesian posterior distribution after the $\alpha$-th datapoint, i.e., after obtaining $\vec{n}_{\alpha}$ as the outcome, is given by
\begin{equation}\label{inference}
    \begin{split}
        P_{\alpha}(\varphi) &= \int_{\bm{\epsilon}}P_{\alpha}(\varphi, \bm{\epsilon}) \,, \\ 
        \text{   where,}\\
        P_{\alpha}(\varphi, \bm{\epsilon}) &= \frac{P(\vec{n}_{\alpha}| \varphi, \bm{\epsilon})P_{\alpha-1}(\varphi)P_{\text{err}}(\bm{\epsilon})}{P_{\alpha-1}(\vec{n}_{\alpha})} \,, \\
        P_{\alpha-1}(\vec{n}_{\alpha}) &= \int_{\bm{\epsilon}, \varphi}P(\vec{n}_{\alpha}| \varphi, \bm{\epsilon})P_{\alpha-1}(\varphi)P_{\text{err}}(\bm{\epsilon}) \,. 
    \end{split}
\end{equation}
Note that we use $P_{\alpha-1}(\varphi)P_{\text{err}}(\bm{\epsilon})$ in the R.H.S. of the second equation and not $P_{\alpha}(\varphi, \bm{\epsilon})$. This is because, while information about $\varphi$ can be updated, $\bm{\epsilon}$ is freshly sampled after each datapoint --- its values in the $\alpha$-th datapoint and the $(\alpha-1)$-th datapoint are uncorrelated random variables, both sampled from $P_{\text{err}}(\bm{\epsilon})$. That is, after receiving each datapoint, we are updating $\varphi$, but learning $\bm{\epsilon}^{(\alpha)}$ afresh. That is, we expect 
\begin{equation}
    \int \bm{\epsilon}P_{\alpha}(\varphi, \bm{\epsilon}) \approx \bm{\epsilon}^{(\alpha)} \,. 
\end{equation}

The precision in the estimate of $\varphi$ after the $\alpha$-th update is given by 
\begin{equation}
    (\Delta^2\varphi)_{\alpha}=\int \varphi^2 P_{\alpha}(\varphi)-\left(\int \varphi P_{\alpha}(\varphi)\right)^2 \,. 
\end{equation}

\subsection{Effective Fisher information}
We use the van Trees bounds to estimate this error. We first define a vector $\bm{x}=(\varphi, \epsilon_1, \cdots, \epsilon_{\ell})^T$ containing all the parameters being estimated in that data point and define the $(\ell+1)\times(\ell+1)$ covariance matrix $\Sigma^{(\alpha)}$:
\begin{equation}
    \Sigma^{(\alpha)}_{ij} = \int x_i x_j P_{\alpha}(\varphi, \bm{\epsilon})-\left(\int x_i P_{\alpha}(\varphi, \bm{\epsilon})\right)\left(\int x_j P_{\alpha}(\varphi, \bm{\epsilon})\right)
\end{equation}
for  $i, j = 1, \cdots, \ell+1$. Note that $\bm{x}=(x_1, \cdots, x_{\ell+1})$. Indeed, $x_1=\varphi$ and therefore, the required error is in the first entry, i.e., $(\Delta^2\varphi)_{\alpha} = \Sigma^{(\alpha)}_{11}$. Moreover, $x_2, \cdots, x_{\ell+1}= \epsilon_1, \cdots, \epsilon_{\ell}$. The van Trees inequality, which is also the Bayesian Cramer-Rao bound reads:
\begin{equation}
\begin{split}
    \Sigma^{(\alpha)}\succeq \left( \bm{I}^{(\alpha-1)}+F\right)^{-1} \,, \\
\end{split}
\end{equation}
where $F$ is the CFI:
\begin{equation}
    F_{ij} = \sum_{\vec{n}} \frac{1}{P(\vec{n}|\varphi, \bm{\epsilon})}\frac{\partial P(\vec{n}|\varphi, \bm{\epsilon})}{\partial x_i} \frac{\partial P(\vec{n}|\varphi, \bm{\epsilon})}{\partial x_j}
\end{equation}
for $i, j = 1, \cdots, \ell+1$ and $\bm{I}^{(\alpha-1)}$ is the prior information defined by:
\begin{equation}
\begin{split}
    &\bm{I}^{(\alpha-1)}_{ij} = \\
    &\int \frac{1}{P_{\alpha-1}(\varphi) P_{\text{err}}(\bm{\epsilon})}\frac{\partial \left( P_{\alpha-1}(\varphi) P_{\text{err}}(\bm{\epsilon})\right)}{\partial x_i} \frac{\partial \left( P_{\alpha-1}(\varphi) P_{\text{err}}(\bm{\epsilon})\right)}{\partial x_j} \,. 
\end{split}
\end{equation}
The prior information can be interpreted as the inverse of the variance. In fact due to the factorization of the prior distribution, $P_{\alpha-1}(\varphi) P_{\text{err}}(\bm{\epsilon})$, which in turn is a consequence of the non-Markovian nature of $\bm{\epsilon}$, $\bm{I}^{(\alpha-1)}$ takes a trivial form. Because $\epsilon_i$ are uncorrelated, $\bm{I}^{(\alpha-1)}$ is a diagonal matrix. Cauchy-Schwarz inequality reads:
\begin{equation}
\begin{split}
    \bm{I}^{(\alpha-1)}_{11} &\geq \frac{1}{ (\Delta^2\varphi)_{\alpha-1}} \,, \\
    \bm{I}^{(\alpha-1)}_{22}&\geq \frac{1}{\sigma_1^2} \,, \\
    &\vdots\\
    \bm{I}^{(\alpha-1)}_{(\ell+1)(\ell+1)}&\geq \frac{1}{\sigma_{\ell}^2} \,. 
\end{split}
\end{equation}
The inequalities saturate when the underlying distribution is Gaussian or Gaussian-like. We may assume that the errors are described by a Gaussian. That is, $P_{\text{err}}$ is a Gaussian. Moreover, the above inequality saturates when the variances approach zero --- which can be applied to $(\Delta^2\varphi)_{\alpha-1}$. Therefore, to a good approximation, 
\begin{equation}
    \bm{I}^{(\alpha-1)}=\text{diag}\left(\frac{1}{ (\Delta^2\varphi)_{\alpha-1}}, \frac{1}{\sigma_1^2}, \cdots, \frac{1}{\sigma_{\ell}^2}\right) \,. 
\end{equation}
Note that while $\bm{I}^{(\alpha)}$ has a simple diagonal form, $F$ in general has significant off-diagonal entries that represent coupling between the errors and the variable $\varphi$. Therefore, the precision $(\Delta^2 \varphi)_{\alpha} = \Sigma^{(\alpha)}_{11}\geq [\left( \bm{I}^{(\alpha-1)}+F\right)^{-1}]_{11} $ can be affected by all terms of $F$. We define the \textit{effective Fisher information}, $F_{\text{eff}}$ accordingly:
\begin{equation}
    \begin{split}
        (\Delta^{2}\varphi)_{\alpha}&\geq \left[\left( \bm{I}^{(\alpha-1)}+F\right)^{-1}\right]_{11} = \frac{1}{F^{(\alpha)}_{\text{eff}}} \,, \\
        F^{(\alpha)}_{\text{eff}}&=\frac{1}{\left[\left( \bm{I}^{(\alpha-1)}+F\right)^{-1}\right]_{11}} \,. 
    \end{split}
\end{equation}
Despite its complicated looking form, it maintains the expected linearity with the number of repetitions $\nu$. Indeed, one can show 
\begin{equation}
    F_{\text{eff}}^{(\alpha)} = \alpha(F_{\text{eff}}^{(1)}-I^{\text{prior}}_{\varphi} ) + I^{\text{prior}}_{\varphi} \,. 
\end{equation}
See the Supplementary Material for a derivation. Here, $ I^{(0)}_{\varphi}$ is the prior on $\varphi$. One can assume $ I^{\text{prior}}_{\varphi}\sim\frac{1}{(\Delta^2\varphi)_{\text{prior}}}$.  In particular, after $\nu$ datapoints, $F_{\text{eff}}^{(\nu)} = \nu(F_{\text{eff}}^{(1)}-I^{\text{prior}}_{\varphi} ) + I^{\text{prior}}_{\varphi}$ is linear in $\nu$ and $F^{(1)}_{\text{eff}}$. It is clear that the latter is the quantity to optimize when designing experiments.

\subsection{Maximal effective Fisher information and Holevo bounds}

For brevity, hereafter we drop the superscript $(1)$ and use $F_{\text{eff}}$ instead. Although it is clear, let us note its definition again:
\begin{equation}
     F_{\text{eff}}=F_{\text{eff}}^{(1)}=\frac{1}{\left[\left( \bm{I}^{\text{prior}}+F\right)^{-1}\right]_{11}} \,. 
\end{equation}
Note that $F$ is the CFI, and depends on the basis. We can therefore maximize the effective Fisher information over $U_M$ -- equivalently, over the PVMs. That is, 

\begin{equation}
    F_{\text{eff, max}} = \max_{\{\tilde{\Pi}_i\}}\frac{1}{\left[\left( \bm{I}^{\text{prior}}+F\right)^{-1}\right]_{11}} \,. 
\end{equation}
This maximum is related to the Holevo bounds. While we have been using the CFI, one of the trivial bounds comes from the quantum Fisher information (QFI), $F_Q$ w.r.t $H_0, H_1, \cdots, H_{\ell}$. It follows that 
\begin{equation}\label{QFIeff}
    F_{\text{eff}}\leq \frac{1}{\left[\left( \bm{I}^{\text{prior}}+F_Q\right)^{-1}\right]_{11}}=F_{\text{eff}}^Q \,. 
\end{equation}
See the Supplementary Material for a proof. This inequality is tight when the Hamiltonians $H_i$ commute with each other and with $H_0$. In the next section, we will use this inequality to study a physically relevant problem.

\section{Correcting spatial inhomogeneity in gravimetry}
In this section, we will study the scaling of $F_{\text{eff}}$ with the number of atoms $N$ for a specific form of error --- spatial inhomogeneity of the potential. Assuming that the lattice is in the $z$ direction, we model the total potential during the $\alpha$-th datapoint as 
\begin{equation}
    V^{(\alpha)}(z) = \eta z + \epsilon^{(\alpha)}_1 f_1(z) + \epsilon^{(\alpha)}_2 f_2(z) +\cdots +  \epsilon^{(\alpha)}_{\ell} f_{\ell}(z)  \,. 
\end{equation}
Here, $\epsilon_i^{(\alpha)}$ are random numbers sampled from independent normal distributions: $\epsilon_i^{(\alpha)} \sim \mathcal N(0, \sigma_i) $. We are assuming that they all have a zero mean and are uncorrelated. Any non-zero mean is a systematic value and can be calibrated out; any correlation can be removed by picking the appropriate linear combinations of $\epsilon_1\cdots\epsilon_{\ell}$. $f_i(z)$ is the inhomogeneity in the potential caused by the $i$-th source. This could be due to additional sources of gravity, with some non-linearity in $z$, laser power variations etc. We assume that the $L$ sites of the lattice are at locations $z_1, \cdots, z_L$ and set $f_{ij}=f_i(z_j)$ so that
\begin{equation}
    H_i = \sum_{j=1}^L f_{ij}\hat{n}_j \,. 
\end{equation}
Assuming that the prior variance on $\varphi$ is $\sigma^2_{\varphi}$, i.e., $\varphi = \eta \tau$ is known up to a precision of $\sigma_{\varphi}$, the prior information matrix is given by $\bm{I}^{\text{prior}} = \text{diag}(1/\sigma_{\varphi}^2, 1/\sigma_1^2, \cdots, 1/\sigma_{\ell}^2)$.

Note that $[H_i, H_j]=0 $ for each pair $j,i$ and they all commute with $H_0$. Therefore, the inequality in Eq.~(\ref{QFIeff}) is tight, making the computation of the maximal $F_{\text{eff}}$ easier. We need to evaluate the $(\ell+1)\times(\ell+1)$ matrix $F_Q$ whose $ij$-th element is given by $[F_Q]_{ij} = \langle H_i H_j\rangle - \langle H_i\rangle \langle H_j\rangle$ for $i, j \in \{0, 1, \cdots, \ell\}$ for a state $\ket{\psi}$. Given the form of $H_i$, $F_Q$ has a complicated form. But it can be simplified using the following observation. Let us define $F_{\mathcal N}$ as the $L\times L$ QFI of the occupancies $\hat{n}_i$, that is, 
$$
[F_{\mathcal N}]_{ij} = \langle \hat{n}_i \hat{n}_j\rangle -\langle \hat{n}_i\rangle \langle \hat{n}_j\rangle \,. 
$$
We then define the $L\times (\ell+1)$ matrix, $\bm{f}$ as, 
\begin{equation}
    \begin{split}
        \bm{f}_{0j} &= j \,, \\
        \bm{f}_{ij} &= f_{ij} \text{ for } i =1, \cdots, \ell \,. 
    \end{split}
\end{equation}
Note that the first row of $\bm{f}$ contains the coefficients of $\hat{n}_i$ in $H_0$. Now, it follows that
\begin{equation}
    F_Q = \bm{f} F_{\mathcal N}\bm{f}^T \,. 
\end{equation}
We are interested in the inverse of $(\bm{I}^{\text{prior}}+ F_Q)$, following Eq.~(\ref{QFIeff}). More precisely, the $_{11}$-th entry of the inverse. Noting that both  $\bm{I}^{\text{prior}}$ and $F_Q$ are positive semi-definite (PSD), a diagonal entry of its inverse is typically a weighted sum of the inverse eigenvalues. That is, if $\lambda_i$ are the eigenvalues of  $(\bm{I}^{\text{prior}}+ F_Q)$, the effective Fisher information is $F_{\text{eff}}=(\sum_i\frac{s_i}{\lambda_i})^{-1}$ where $s_i\geq 0$ and $\sum s_i=1$. Indeed $s_i$ are the squares of the coefficients of the first vector (labeled $`1'$ in the matrix form) in the eigenbasis of $(\bm{I}^{\text{prior}}+ F_Q)$. We will use elementary inequalities to find bounds on $F_{\text{eff.}}$ in terms of $\sigma_i^2$ and the eigenvalues of $F_Q$, $F_1, \cdots, F_{\ell+1}$. Note that $1/\sigma_i^2$ are the eigenvalues of $\bm{I}^{\text{prior}}$. It follows that the minimum and maximum eigenvalues of $(\bm{I}^{\text{prior}}+ F_Q)$ satisfy
\begin{equation}
    \begin{split}
        \lambda_{\min}&\leq \frac{1}{\sigma_{\min}^2}+F_{\min} \,, \\
        \lambda_{\max}&\geq \frac{1}{\sigma_{\max}^2}+F_{\max} \,. 
    \end{split}
\end{equation}
See the Supplementary Material for a proof. That is, there is at least one eigenvalue larger than $\frac{1}{\sigma_{\max}^2}+F_{\max}$ and at least one smaller than $\frac{1}{\sigma_{\min}^2}+F_{\min}$. That sets the range of the $\frac{1}{\lambda_i}'s$ of which $F_{\text{eff}}$ is a linear combination. We assume that all of the priors, $\sigma_i^2$ are within $\mathcal O(1)$ --- they do not vary much. We therefore replace $\sigma_{\max}$ and $\sigma_{\min}$ by $\sigma_{\text{err}}$, a representative standard deviation of the errors. This results in
\begin{equation}
    \Delta^2\varphi \geq F_{\text{eff}}^{-1} = \frac{a \sigma_{\text{err}}^2}{1+F_{\min}\sigma_{\text{err}}^2} +  \frac{b\sigma_{\text{err}}^2}{1+F_{\max}\sigma_{\text{err}}^2} \,, 
\end{equation}
where $a, b$ are $\mathcal O(1)$ constants. We expect the eigenvalues of $F_Q$ to scale as $\mathcal O(N^2)$. If both $F_{\min}$ and $F_{\max}$ scale as $\mathcal O(N^2)$, so does $F_{\text{eff}}$. A particularly interesting case is when $F_Q$ is singular --- $F_{\min}=0$. Even if $F_{\max}\sim \mathcal O(N^2)$, the effective Fisher information saturates at $\mathcal O (1/\sigma_{\text{err}}^2)$. Therefore, it is important to note the rank of $F_Q$. Following $\sum_j \hat{n}_j=N$, the rank of $F_{\mathcal N}$ is $L-1$. The rank of $\bm{f}$ is the minimum of $(\ell+1)$ and $L$. Therefore, 
\begin{equation}
    \text{rank}(F_Q) = \min\{L-1, \ell+1\} \,. 
\end{equation}
Thus, $F_Q$ is not a full rank matrix when $L<\ell+2$. Hence, we have the following result:
\begin{equation}
    F_{\text{eff}} \leq \frac{\mathcal{O}(1)}{\sigma_{\text{err}}^2} \text{ if } L<\ell+2 \,. 
\end{equation}
This allows for no Heisenberg (or even SQL) scaling. We will next show that when $L\geq \ell+2$, there are easily accessible states for which
\begin{equation}\label{F_eff_HL}
    F_{\text{eff}} \geq \mathcal{O}(N^2) \text{ if } L\geq \ell+2 \,. 
\end{equation}
To see this, let us consider the typical values of $F_{\mathcal N}$ in the Hilbert space. We derive the following results in the Supplementary Material:
\begin{equation}\label{Haar_avg}
    \begin{split}
        \int_{\text{Haar}} \langle \hat{n}_i^2\rangle -\langle \hat{n}_i\rangle ^2  &= \frac{N(N+L)(L-1)}{L^2(L+1)} \,, \\
        \int_{\text{Haar}} \langle \hat{n}_i\hat{n}_j\rangle -\langle \hat{n}_i\rangle \langle \hat{n}_j\rangle   &= -\frac{N(N+L)}{L^2(L+1)} \,. \\
    \end{split}
\end{equation}
Moreover, the width of the distributions around these typical values is vanishingly small --- it scales as $\frac{1}{\sqrt{d}}$, where $d=\binom{N+L-1}{N}\approx \frac{N^{L-1}}{(L-1)!}$ is the Hilbert space dimension. 
\begin{equation}\label{Variance}
\begin{split}
    \text{Var}[\hat{n}_i^2\rangle -\langle \hat{n}_i\rangle ^2] &\approx \frac{N^4}{d} \,, \\
    \text{Var}[\langle \hat{n}_i\hat{n}_j\rangle -\langle \hat{n}_i\rangle \langle \hat{n}_j\rangle] &\approx \frac{N^4}{d} \,. \\
\end{split}
\end{equation}
Note that while these averages are calculated for the Haar measure, they only use upto the $4$-th moments and are valid for any $4-$design. Therefore, the $F_{\mathcal N}$ for \textit{almost} any state sampled from \textit{any} $4-$design would be 
\begin{equation}\label{F_4D_matrix}
\begin{split}
        F_{\mathcal N} &= \frac{N(N+L)}{L(L+1)}\begin{pmatrix}
1 & 0 & \cdots & 0\\
0 & 1 & \cdots & 0\\
\vdots & \vdots & \ddots & \vdots\\
0 & 0 & \cdots & 1
\end{pmatrix}- \frac{N(N+L)}{L^2(L+1)}\begin{pmatrix}
1 & 1 & \cdots & 1\\
1 & 1 & \cdots & 1\\
\vdots & \vdots & \ddots & \vdots\\
1 & 1 & \cdots & 1
\end{pmatrix} \,. 
\end{split}
\end{equation}
$L-1$ of the eigenvalues are $\frac{N(N+L)}{L(L+1)}$ and the last eigenvalue is $0$. Understandably, the eigenvector with zero eigenvalue is $(1,1,\cdots, 1)$ --- because $\sum_i \hat{n}_i=N$. We will assume, without any loss of generality that each of the row vectors in $\bm{f}$ are orthogonal to $(1,1,\cdots, 1)$. This is equivalent to saying $\text{Tr}[H_i]=0~\forall~i$. We can now write $F_Q$ in a very simple form:
\begin{equation}
    F_Q = \frac{N(N+L)}{L(L+1)} \bm{f}\bm{f}^T \,. 
\end{equation}
Therefore, unless $\bm{f}\bm{f}^T$ is singular, each eigenvalue of $F_Q$ is $\mathcal O\left(\frac{N^2}{L^2}\right)$. When $L\geq \ell+2$ it now follows that for most states, $F_{\text{eff}}$ scales as $\frac{N^2}{L^2}$. 

\begin{figure}
    \centering
    \includegraphics[width=0.99\linewidth]{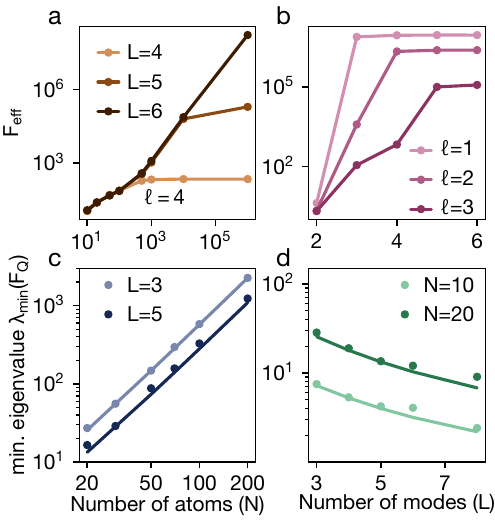}
    \caption{\textbf{Numerical illustrations:} \textbf{a.} The effective Fisher information $F_{\text{eff}}$ for $\ell=4$ channels of error saturates for any $L<\ell+2=6$ due to a singularity in the QFI. For $L\geq \ell+2$, it scales asymptotically as $\mathcal O(N^2)$. \textbf{b.} $F_{\text{eff}}$ at a fixed $N=10^4$, showing that one needs at least $L=\ell+2$ to mitigate the errors. Each curve corresponds to a different $\ell$ and saturates at $L=\ell+2$.  \textbf{c, d.} Minimum eigenvalue of the $F_Q$ of a state obtained after a random experimental sequence. The markers are numerical results and the solid line is the formula in Eq.~(\ref{typ}). This shows that we can generate a random pulse sequence on the fly and produce a Heisenberg scaling in error mitigated quantum sensing of gravity.}
    \label{Fig3}
\end{figure}

In Fig.~\ref{Fig3}a, b we show numerically computed $F_{\text{eff}}$ for a few examples. We use $\sigma_{\varphi}=\sigma_i = 0.01$ for all $i=1, \cdots, \ell$ and generate random errors where each $f_{ij}$ is uniformly random in $[-1,1]$.  Fig.~\ref{Fig3}a clearly shows the asymptotic behavior of $F_{\text{eff}}$ ---  it saturates for $L<\ell+2$ and scales as $N^2$ for $L\geq \ell+2$.  Fig.~\ref{Fig3}b shows how the effective Fisher information improves as we increase $L$, until it crosses $\ell+2$, after which it saturates. In the next section, we discuss a potential experimental implementation. 

\section{Proposed experiment}
The typical value of $F_{\mathcal N}$, shown in Eq.~(\ref{Haar_avg}) and Eq.~(\ref{Variance}) suggests that one can use a random experimental control sequence to produce a metrologically useful state. That is $U_{SP}$ can be produced by a sequence of \textit{random} control pulses. In ref.~\cite{shao2025engineeringmultimodebosonicsqueezed}, we used this idea to develop a Monte-Carlo optimization technique. Here, we argue that one does not even need to optimize $U_{SP}$ --- almost any $U_{SP}$ can be used. We develop a  Loschmidt echo experimental sequence using this idea.

\subsection{Loschmidt echo sequence}
The idea is pick a random experimental control sequence to produce $U_{SP}$ and echo the sequence to produce $U_M =U_{SP}^{-1}$. This ensures that the measurement in the $\{\hat{n}_i\}$ basis saturates the CFI-QFI inequality, i.e., $F=F_Q$. 
To develop the random control pulse sequence, we will use the structure developed in ref.\cite{shao2025engineeringmultimodebosonicsqueezed} and start with an initial state, $\ket{\psi_0}=(\hat{a}_1^{\dagger})^N/\sqrt{N!}\ket{\text{vac}}$, i.e., all atoms in the first lattice site. We evolve it up to time $T$ (here, $T$ is the time taken for the state preparation), by applying $2n$ evenly paced experimental pulses, for an integer $n$, where the width of each pulse is $\Delta t = T/(2n)$. Note that the Hamiltonian is time-independent within each pulse $\Delta t$. Further, we evolve/apply three Hamiltonians below sequentially. At the $k$-th step,

\begin{eqnarray}\label{H1H2H3}
    \begin{split}
        H_1^{(k)}&=\sum_i \Delta_i^{(k)} \hat{n}_i + U_i \hat{n}_i(\hat{n}_i-1) \,, \\
        H_2^{(k)}&=\sum_i J_{i}^{(k)} (\hat{a}^{\dagger}_{i+1}\hat{a}_{i}+\hat{a}^{\dagger}_{i}\hat{a}_{i+1}) \,. \\
    \end{split}
\end{eqnarray}
We pick $3$ random matrices --- a $n\times (L-1) $ matrix $J^{(k)}_i$ (it can be treated as a matrix because $k=1,\cdots, n$ and $i=1, \cdots, L-1$), and $n\times L$ matrices $\Delta^{(k)}_i$ and $U^{(k)}_i$. We pick the matrix $J^{(k)}_i$ with uniformly random entries from $[0,1]$, and matrices $\Delta^{(k)}_i$ and $U^{(k)}_i$ with uniformly random entries from $[-1,1]$. We then apply the pulse sequences accordingly to implement the following state preparation unitary:
\begin{equation}\label{eq:final}
U_{SP} = \Pi_ke^{-i\Delta t H_2^{(k)}}e^{-i\Delta t H_1^{(k)}} \ . 
\end{equation}
The final state would be $U_{SP}\ket{\psi_0}$. For a random choice of $J^{(k)}_i, \Delta^{(k)}_i$ and   $U^{(k)}_i$, we will almost certainly produce a state with $F_{\text{eff}}=\mathcal O (N^2/L^2)$ ~\cite{shao2025engineeringmultimodebosonicsqueezed}.  Indeed, the typical value of the minimum eigenvalue of $F_Q$, $F_{\min}$ is,
\begin{equation}\label{typ}
    F_{\min} = \frac{N(N+L)}{L(L+1)} \,. 
\end{equation}
 In Fig.~\ref{Fig3}c,d, we show that a random experimental pulse sequence indeed consistently produces a state with $F_{\text{eff}}$. In Fig.~\ref{Fig3}c, we show that the minimum eigenvalue of $F_Q$ after a random pulse sequence indeed follows Eq.~(\ref{typ}) and therefore the effective Fisher information would scale as $N^2$. In Fig.~\ref{Fig3}d, we demonstrate the same, for larger values of $L$.   In ref. ~\cite{shao2025engineeringmultimodebosonicsqueezed},  we have provided a mathematical proof that the states produced by random evolutions under Hamiltonians of the form Eq.~(\ref{H1H2H3}) approach a $4-$design and therefore,  states produced under this evolution almost certainly have a QFI given by Eq.~(\ref{F_4D_matrix}). Accordingly, for these states,  $F_{\text{eff}}=\mathcal O (N^2/L^2)$.  

\begin{center}
\fbox{
\parbox{0.95\linewidth}{

\textbf{Loschmidt echo sequence:}\\
Repeat for $\alpha=1, \cdots, \nu$
\begin{enumerate}
    \item Prepare the atoms in the state
    \[
        \ket{\psi_0} = \frac{(\hat{a}_1^{\dagger})^N}{\sqrt{N!}}\ket{\text{vac}}.
    \]
    \item Divide the total time interval $[0,T]$ into $2n$ uniform sections of size $\Delta t = T/(2n)$.

    \item Generate random matrices:
        \[
        \begin{aligned}
        J^{(k)}_i &\in [0,1], &i=1,\dots,L-1,\\
        \Delta^{(k)}_i, U^{(k)}_i &\in [-1,1], &i=1,\dots,L,
        \end{aligned}
        \]
        for $k = 1, \dots, n$.
        
        \item Apply the pulse sequence:
        \begin{equation*}
            \Delta_i(t) = \begin{cases}
                \Delta_i^{(k)} \text{ if   }\  (2k-2)\Delta t \leq t \leq (2k-1)\Delta t\\
                0 \text{  otherwise}
            \end{cases}
        \end{equation*}
        \begin{equation*}
            U_i(t) = \begin{cases}
                U_i^{(k)} \text{ if   }\  (2k-2)\Delta t \leq t \leq (2k-1)\Delta t\\
                0 \text{  otherwise}
            \end{cases}
        \end{equation*}
        \begin{equation*}
            J_i(t) = \begin{cases}
                J_i^{(k)} \text{ if   }\  (2k-1)\Delta t \leq t \leq 2k\Delta t\\
                0 \text{  otherwise}
            \end{cases}
        \end{equation*}

    \item Turn off all control parameters ($\Delta_i, U_i, J_i=0$) and allow for phase acquisition for a duration $\tau$.
    \item Apply the same pulse sequence in reverse order, to implement an echo. 
    \item Readout in the $\{\hat{n}_i\}$ basis to obtain a datapoint $\vec{n}_{\alpha} = (n_1, \cdots, n_L)$. 
    \item Use Eq.~(\ref{inference}) to update the value of the unknown $\varphi$. 
\end{enumerate}

}}

\vspace{2mm}

{ Box 1: Proposed experimental  sequence}\label{Algorithm}
\end{center}

\section{Relation to non-Hermitian dynamics}
We discuss mathematical formalizations of the earlier comment we made on how the ``extra" degrees of freedom in an $L$-mode system are being used to detect and post-correct errors. Let us simplify the system by setting $N=1$ atom in a lattice with $L$ sites -- we are now in an $L$ dimensional Hilbert space. The local sensitivity of a single atom state $\ket{\psi}$ in this Hilbert space to $\varphi$ is given by the distance between $\ket{\psi}$ and $e^{-i\varphi H_0}\ket{\psi}$ when $\varphi$ is infinitesimal. The local dynamics, i.e., when $\varphi$ is infinitesimal occurs in a $2-$dimensional subspace of the $L$ dimensional Hilbert space spanned by $\ket{\psi}$ and $H_0\ket{\psi}$.  In the absence of errors, the local dynamics remains in the 2-dimensional space. However, it leaks into the remaining $L-2$ dimensional subspace in the presence of errors generated for instance, by $H_1, \cdots$. Therefore, one can formulate the ideas presented in this paper also as measurement on the leakage space followed by post-correction.

One of the fruitful ways to formulate measurement-feedback on a ``leakage" space is non-Hermitian dynamics. In refs.~\cite{Naghiloo2021}, two-level non-Hermitian dynamics was implemented using a third ``leakage" level. Population in this level was used to implement post-selected dynamics, which is effectively described by a non-Hermitian Hamiltonian. The post-correction developed here can be formulated as a non-Hermitian dynamics of an effective 2-level system described by the span of $\ket{\psi}$ and $H_0\ket{\psi}$. The effective Fisher information $F_{\text{eff}}$ can be connected to the quantum Fisher information in non-Hermitian dynamics~\cite{Naikooetal2025, Lietal2025}. Interestingly, better-than-SQL precision has been shown extensively in non-Hermitian systems~\cite{McDonaldetal2020, Yuetal2024}. Using a non-Hermitian dynamics formulation  also allows us to expand the scope of this work to errors originating from open system dynamics, which can be classically detected by monitoring photon emissions for instance.

\section{Conclusions}
We have developed a technique for error mitigation in quantum sensing of gravity with multi-mode Bose-Einstein condensates. We address non-Markovian errors generated by random, fluctuating error Hamiltonians acting alongside gravity.  We have shown that these errors can be mitigated using a Bayesian post-correction if the number of modes ($L$) and the number of independent sources of error ($\ell$) satisfy $L\geq \ell +2$. We also show that under this condition, the post-corrected precision has a Heisenberg scaling in the number of atoms --- we can not only correct the errors,  but we can also recover most of the quantum advantage using our technique. We developed a Loschmidt echo-based protocol to implement our technique experimentally.

A logical next step would be to explore experimental techniques to characterize the  error distribution $P_{\text{err}}(\bm{\epsilon})$ in a given system. This can benefit from the recently developed benchmarking and error characterization tools~\cite{PhysRevResearch.6.043127, Nielsen_2021}. Exploring  nature of errors $\mathcal E$ against which these techniques are effective would be another important direction~\cite{madhusudhana2025optimizinglossystatepreparation}. 

Yet another important open problem is to extend these results to Markovian errors, coming from open system dynamics.  Arguably, the most important open system effect is loss of atoms in the context of Bose-Einstein condensates. Recent work~\cite{HEBBEMADHUSUDHANA2025100046} has shown that while loss of atoms renders a Heisenberg scaling impossible, in some cases, it allows for a significant sub-SQL scaling offering a scalable quantum advantage.  Developing a Bayesian post-correction protocol to exploit this advantage will be an important direction both for theoretical and experimental work. 

Finally,  the connection to non-Hermitian dynamics is another rich direction.  The broader connection between error correction and non-unitary quantum maps has been the subject of intense investigation~\cite{PhysRevX.10.041020, PhysRevLett.125.030505,kuji2026quantumerrormitigationsimulates}.  Moreover, applications of non-Hermitian physics to quantum sensing is also a hot topic~\cite{PhysRevA.109.062611}.  The relation between our results and non-Hermitian dynamics connects these two paradigms and is therefore a promising direction. 



\section*{Acknowledgements}
Research presented in this article was supported by the Laboratory Directed Research and Development program of Los Alamos National Laboratory under project number 20230779PRD1.  This work was carried out under the auspices of the US Department of Energy NNSA under Contract No. 89233218CNA000001.

\paragraph*{\textbf{Competing interests}} The authors declare no competing interests. 
\bibliography{references}

\cleardoublepage

\setcounter{figure}{0}
\setcounter{page}{1}
\setcounter{equation}{0}
\setcounter{section}{0}

\renewcommand{\thepage}{S\arabic{page}}
\renewcommand{\thesection}{S\arabic{section}}
\renewcommand{\theequation}{S\arabic{equation}}
\renewcommand{\thefigure}{S\arabic{figure}}
\onecolumngrid
\begin{center}
\huge{Supplementary Information}
\vspace{5mm}
\end{center}
\twocolumngrid
\normalsize

\section{Scaling of the effective fisher information (Eq.~(17) of the main text)}
In this section, we will derive Eq.~(17) of the main text and show the linear scaling of the effective Fisher information in the number of statistical repetitions $\nu$. 

The simplest way to solve this problem is to incorporate the fact that the errors $\bm{\epsilon}^{(\alpha)}$ are independent variables for different $\alpha$. That is, after $\alpha$ repetitions of the experiment, the set of outcomes $(\vec{n}_1, \cdots, \vec{n}_{\alpha})$ are a sample from the following distribution:
\begin{equation}
\begin{split}
    P&(\vec{n}_1, \vec{n}_1, \cdots, \vec{n}_{\alpha}|\varphi, \bm{\epsilon}^{(1)}, \bm{\epsilon}^{(2)}, \cdots, \bm{\epsilon}^{(\alpha)})  \\&=P(\vec{n}_1|\varphi,\bm{\epsilon}^{(1)} )P(\vec{n}_2|\varphi,\bm{\epsilon}^{(2)} )\cdots P(\vec{n}_{\alpha}|\varphi,\bm{\epsilon}^{(\alpha)} )
\end{split}
\end{equation}
Note that we now have $(1+\alpha \ell)$ variables. We can define the fluctuations as a $(1+\alpha \ell )\times (1+\alpha \ell)$ $\Sigma-$matrix and the prior $\bm{I}$, also a $(1+\alpha \ell )\times (1+\alpha \ell)$ diagonal matrix. 
\begin{equation}
\begin{split}
    \bm{I} = \text{diag}(I_{\varphi}^{\text{prior}}, 1/\sigma_1^2, \cdots, 1/\sigma_{\ell}^2, 1/\sigma_1^2, \cdots, 1/\sigma_{\ell}^2, \cdots,\\ 1/\sigma_1^2, \cdots, 1/\sigma_{\ell}^2)
\end{split}
\end{equation}
The Van-Tree inequality reads:
\begin{equation}
    \Sigma \succeq (\bm{I}+F^{(\alpha)})^{-1} \implies \Sigma_{11}\geq [(\bm{I}+F^{(\alpha)})^{-1}]_{11}
\end{equation}
Here, $F^{(\alpha)}$ is the $(1+\alpha \ell)\times(1+\alpha \ell)$ CFI matrix. The effective fisher information is given by 
\begin{equation}
    F_{\text{eff}}^{(\alpha)} = \frac{1}{[(\bm{I}+F^{(\alpha)})^{-1}]_{11}} 
\end{equation}

Given that $\bm{\epsilon}^{(i)}$ and $\bm{\epsilon}^{(j)}$ are uncorrelated for $i\neq j$, $F^{(\alpha)} + \bm{I}$ is an \textit{arrowhead matrix}. In particular,
\begin{equation}
    F^{(1)}= \left(
    \begin{array}{ccccc}
        F_{\varphi\varphi} & F_{\varphi \epsilon_1}&\cdots&\cdots &F_{\varphi \epsilon_{\ell}} \\
         F_{\varphi \epsilon_1}&F_{\epsilon_1\epsilon_1} &0&0 & 0\\
         \vdots & 0 & \ddots & 0&0\\
         \vdots & 0& 0&\ddots&0\\
         
         F_{\varphi \epsilon_{\ell}}&0 &0&0 &F_{\epsilon_{\ell}\epsilon_{\ell}} \\
    \end{array}
    \right) = \left(\begin{array}{cc}
        F_{\varphi\varphi} & \bm{v}^T \\
        \bm{v} & F_{\bm{\epsilon}}
    \end{array}\right)
\end{equation}
Here, $F_{\varphi\varphi} = \sum (\frac{\partial}{\partial \varphi}\log (P(\vec{n}|\varphi,\bm{\epsilon} )))^2P(\vec{n}|\varphi,\bm{\epsilon} )$ and $F_{\varphi\epsilon_i} = \sum (\frac{\partial}{\partial \varphi}\log (P(\vec{n}|\varphi,\bm{\epsilon} )))(\frac{\partial}{\partial \epsilon_i}\log (P(\vec{n}|\varphi,\bm{\epsilon} )))P(\vec{n}|\varphi,\bm{\epsilon} )$. Because $\bm{\epsilon}^{(1)}, \bm{\epsilon}^{(2)}\cdots$ are uncorrelated, $F^{(\alpha)}$ has a simple relation to $F^{(1)}$. 

\begin{equation}
    F^{(\alpha)} = \left(\begin{array}{ccccc}
        \alpha F_{\varphi\varphi} & \bm{v}^T& \bm{v}^T&\cdots &\bm{v}^T\\
        \bm{v} & F_{\bm{\epsilon}} & 0 & \cdots &0\\
        \vdots & 0 & \ddots & \cdots &0\\
        \vdots & 0 & 0 & \ddots &0\\
        \bm{v} & 0 & 0 & \cdots &F_{\bm{\epsilon}}\\
    \end{array}\right)
\end{equation}

The inverse of an arrowhead matrix can be computed using Schur complement. It follows that 
\begin{equation}
\begin{split}
    F_{\text{eff}}^{(\alpha)} =\frac{1}{[(\bm{I}+F^{(\alpha)})^{-1}]_{11}} = (\bm{I}+F^{(\alpha)})_{00}-\sum_{k=1}^{\alpha\ell}\frac{(\bm{I}+F^{(\alpha)})^2_{0k}}{(\bm{I}+F^{(\alpha)})_{kk}}\\
    =I^{\text{prior}}_{\varphi}+\alpha F^{(1)}_{00} -\alpha \sum_{k=1}^{\ell}\frac{(\bm{I}+F^{(1)})^2_{0k}}{(\bm{I}+F^{(1)})_{kk}}
\end{split}
\end{equation}
Note that $\bm{I}_{00}=I^{\text{prior}}_{\varphi}$ and $F^{(\alpha)}_{00}=\alpha F^{(1)}_{00}$. It now follows that
\begin{equation}
    F_{\text{eff}}^{(\alpha)} = I^{\text{prior}}_{\varphi} + \alpha (F_{\text{eff}}^{(1)} - I^{\text{prior}}_{\varphi})
\end{equation}

\section{Haar averages: derivation of Eqs.~30, 31 of the main text}
In this section, we provide a derivation of Eqs.~30 and 31 of the main text. Due to symmetry,  we can use $i=1$ and $j=2$ in Eqs.~30 \& 31 of the main text. Let
$$
A=\langle \hat{n}_1^2\rangle-\langle \hat{n}_1\rangle^2
$$
and
$$
B=\langle \hat{n}_1 \hat{n}_2\rangle-\langle \hat{n}_1\rangle\langle \hat{n}_2\rangle
$$
Let $\mathbb{E}[\cdot]$ denote the average over the Haar measure on the unitary group $U(D)$. The integration formulas for the first and second moments of expectation values are:

\begin{align}
    \mathbb{E}[\langle X \rangle] &= \frac{\text{Tr}(X)}{D}, \\
    \mathbb{E}[\langle X \rangle \langle Y \rangle] &= \frac{\text{Tr}(X)\text{Tr}(Y) + \text{Tr}(XY)}{D(D+1)}.
\end{align}

We define the ``trace average'' (or microcanonical average) of an operator as $\overline{X} \equiv \frac{\text{Tr}(X)}{D}$. The Haar average of the covariance between two observables $X$ and $Y$ is:
\begin{align}
    \mathbb{E}[\text{Cov}(X,Y)] &= \mathbb{E}[\langle XY \rangle] - \mathbb{E}[\langle X \rangle \langle Y \rangle] \nonumber \\
    &= \overline{XY} - \frac{\text{Tr}(X)\text{Tr}(Y) + \text{Tr}(XY)}{D(D+1)} \nonumber \\
    &= \frac{D(D+1)\overline{XY} - (D^2 \overline{X} \overline{Y} + D \overline{XY})}{D(D+1)} \nonumber \\
    &= \frac{D^2 \overline{XY} - D^2 \overline{X}\overline{Y}}{D(D+1)} \nonumber \\
    &= \frac{D}{D+1} \left( \overline{XY} - \overline{X} \cdot \overline{Y} \right).
\end{align}
Applying this to our specific quantities:
\begin{align}
    \mathbb{E}[A] &= \frac{D}{D+1} \left( \overline{n_1^2} - (\overline{n_1})^2 \right), \label{eq:HaarA}\\
    \mathbb{E}[B] &= \frac{D}{D+1} \left( \overline{n_1 n_2} - \overline{n_1} \cdot \overline{n_2} \right). \label{eq:HaarB}
\end{align}
We calculate the traces over the basis states $|\mathbf{n}\rangle$.

\subsection{First Moments}
By symmetry among the $L$ modes:
\begin{equation}
    \overline{n_1} = \frac{1}{L} \overline{\sum_{i=1}^L n_i} = \frac{1}{L} \overline{N} = \frac{N}{L}.
\end{equation}
Thus, $\overline{n_2} = \frac{N}{L}$ as well.

\subsection{Second Moments}
We utilize the moments of the Bose-Einstein distribution (or combinatorial falling factorials). The average of $n_i(n_i-1)$ over all states is known to be:
\begin{equation}
    \overline{n_1(n_1-1)} = \frac{N(N-1)}{L(L+1)} \times 2.
\end{equation}
Therefore, the second moment for a single mode is:
\begin{equation}
    \overline{n_1^2} = \overline{n_1(n_1-1)} + \overline{n_1} = \frac{2N(N-1)}{L(L+1)} + \frac{N}{L}.
\end{equation}

For the cross term, we use the identity $N^2 = (\sum n_i)^2 = \sum n_i^2 + \sum_{i \neq j} n_i n_j$. Taking the trace average:
\begin{equation}
    N^2 = L \overline{n_1^2} + L(L-1) \overline{n_1 n_2}.
\end{equation}
Substituting $\overline{n_1^2}$:
\begin{equation}
    N^2 = L \left( \frac{2N(N-1)}{L(L+1)} + \frac{N}{L} \right) + L(L-1) \overline{n_1 n_2}.
\end{equation}
Solving for $\overline{n_1 n_2}$ yields:
\begin{equation}
    \overline{n_1 n_2} = \frac{N(N-1)}{L(L+1)}.
\end{equation}

\subsection{Calculating $\mathbb{E}[A]$}
Substituting the moments into Eq. (\ref{eq:HaarA}):
\begin{align*}
    \Delta_A &= \overline{n_1^2} - (\overline{n_1})^2 \\
    &= \frac{2N(N-1)}{L(L+1)} + \frac{N}{L} - \frac{N^2}{L^2} \\
    &= \frac{N}{L} \left[ \frac{2L(N-1) + L(L+1) - N(L+1)}{L(L+1)} \right] \\
    &= \frac{N}{L^2(L+1)} \left[ 2LN - 2L + L^2 + L - NL - N \right] \\
    &= \frac{N}{L^2(L+1)} \left[ LN + L^2 - L - N \right] \\
    &= \frac{N(L-1)(N+L)}{L^2(L+1)}.
\end{align*}
Applying the Haar scaling factor:
\begin{equation}
   \mathbb{E}[A] = \frac{D}{D+1} \frac{N(N+L)(L-1)}{L^2(L+1)}
\end{equation}

\subsection{Calculating $\mathbb{E}[B]$}
Substituting the moments into Eq. (\ref{eq:HaarB}):
\begin{align*}
    \Delta_B &= \overline{n_1 n_2} - \overline{n_1}\overline{n_2} \\
    &= \frac{N(N-1)}{L(L+1)} - \frac{N^2}{L^2} \\
    &= \frac{N}{L} \left[ \frac{L(N-1) - N(L+1)}{L(L+1)} \right] \\
    &= \frac{N}{L^2(L+1)} \left[ LN - L - NL - N \right] \\
    &= -\frac{N(N+L)}{L^2(L+1)}.
\end{align*}
Applying the Haar scaling factor:
\begin{equation}
\mathbb{E}[B] = -\frac{D}{D+1} \frac{N(N+L)}{L^2(L+1)}
\end{equation}
\subsection{Approximation of the second moments, Eq.~31 of the main text}
We define the fluctuation of an operator's expectation value from its Hilbert space average as:
\begin{equation}
    \delta \langle \mathcal{O} \rangle \equiv \langle \psi | \mathcal{O} | \psi \rangle - \overline{\mathcal{O}}, \quad \text{where} \quad \overline{\mathcal{O}} = \frac{\text{Tr}(\mathcal{O})}{D}.
\end{equation}
By definition, $\mathbb{E}[\delta \langle \mathcal{O} \rangle] = 0$. The Haar integral formula for the second moment provides the correlation between fluctuations:
\begin{equation}
    \mathbb{E}[\delta \langle X \rangle \delta \langle Y \rangle] = \frac{D \overline{XY} - D \overline{X}\cdot\overline{Y}}{D(D+1)} \approx \frac{1}{D} \left( \overline{XY} - \overline{X}\cdot\overline{Y} \right) = \frac{1}{D} \text{Cov}_T(X, Y).
\end{equation}
The variance of a quantity $f(\langle X \rangle, \langle Y \rangle)$ is $\mathbb{E}[f^2] - (\mathbb{E}[f])^2$. To find the leading order scaling, we expand $f$ in powers of $\delta$.
\begin{equation}
    f \approx f(\overline{X}, \overline{Y}) + \sum_i \frac{\partial f}{\partial \langle X_i \rangle} \delta \langle X_i \rangle + \frac{1}{2} \sum_{i,j} \frac{\partial^2 f}{\partial \langle X_i \rangle \partial \langle X_j \rangle} \delta \langle X_i \rangle \delta \langle X_j \rangle.
\end{equation}
The variance is dominated by the square of the linear term:
\begin{equation}
    \text{Var}(f) \approx \mathbb{E}\left[ \left( \sum_i \frac{\partial f}{\partial \langle X_i \rangle} \delta \langle X_i \rangle \right)^2 \right] \sim O(1/D).
\end{equation}
Higher order terms (involving products of 3 or 4 $\delta$'s) scale as $O(1/D^2)$ or higher.

\subsection{Trace Moments in the Limit $N, L \gg 1$}

We consider the limit of large particle number and large mode number. Let $\lambda = N/L$ be the filling factor. The trace moments correspond to the moments of the Bose-Einstein distribution. For large $N$, $\overline{n^k} \approx k! \lambda^k$.

The required trace statistics are:
\begin{align}
    \overline{n_1} &\approx \lambda, & \text{Var}_T(n_1) &\approx 2\lambda^2 - \lambda^2 = \lambda^2. \\
    \overline{n_1^2} &\approx 2\lambda^2, & \text{Var}_T(n_1^2) &\approx 24\lambda^4 - (2\lambda^2)^2 = 20\lambda^4. \\
    \overline{n_1 n_2} &\approx \lambda^2, & \text{Var}_T(n_1 n_2) &\approx 4\lambda^4 - (\lambda^2)^2 = 3\lambda^4.
\end{align}
We also require the covariance between the squared number and number operator:
\begin{equation}
    \text{Cov}_T(n_1^2, n_1) = \overline{n_1^3} - \overline{n_1^2}\cdot\overline{n_1} \approx 6\lambda^3 - (2\lambda^2)(\lambda) = 4\lambda^3.
\end{equation}
And the covariance between cross-terms and single modes:
\begin{equation}
    \text{Cov}_T(n_1 n_2, n_1) = \overline{n_1^2 n_2} - \overline{n_1 n_2}\cdot\overline{n_1} \approx 2\lambda^3 - \lambda^3 = \lambda^3.
\end{equation}

\subsection{Variance of A}
We define $A = \langle n_1^2 \rangle - \langle n_1 \rangle^2$. Let $Q_1 = n_1^2$ and $Q_2 = n_1$.
Expanding around the means:
\begin{align}
    A &= (\overline{Q_1} + \delta \langle Q_1 \rangle) - (\overline{Q_2} + \delta \langle Q_2 \rangle)^2 \nonumber \\
    &= (\overline{Q_1} - \overline{Q_2}^2) + \underbrace{\delta \langle Q_1 \rangle - 2\overline{Q_2} \delta \langle Q_2 \rangle}_{\text{Linear Term}} - \underbrace{(\delta \langle Q_2 \rangle)^2}_{\text{Quad Term}}.
\end{align}
The variance is given to leading order by the expectation of the square of the linear term:
\begin{align}
    \text{Var}(A) &\approx \mathbb{E}\left[ (\delta \langle Q_1 \rangle - 2\overline{Q_2} \delta \langle Q_2 \rangle)^2 \right] \nonumber \\
    &= \mathbb{E}[ (\delta \langle Q_1 \rangle)^2 ] - 4\overline{Q_2} \mathbb{E}[ \delta \langle Q_1 \rangle \delta \langle Q_2 \rangle ] + 4\overline{Q_2}^2 \mathbb{E}[ (\delta \langle Q_2 \rangle)^2 ] \nonumber \\
    &= \frac{1}{D} \left[ \text{Var}_T(Q_1) - 4\overline{Q_2} \text{Cov}_T(Q_1, Q_2) + 4\overline{Q_2}^2 \text{Var}_T(Q_2) \right].
\end{align}
Substituting the moments from Section 2:
\begin{align}
    \text{Var}(A) &\approx \frac{1}{D} \left[ 20\lambda^4 - 4(\lambda)(4\lambda^3) + 4(\lambda)^2(\lambda^2) \right] \nonumber \\
    &= \frac{1}{D} \left[ 20\lambda^4 - 16\lambda^4 + 4\lambda^4 \right] \nonumber \\
    &= \frac{8\lambda^4}{D}.
\end{align}
\begin{equation}
    \text{Var}(A) \approx \frac{8}{D} \left( \frac{N}{L} \right)^4
\end{equation}

\subsection{Variance of B}
We define $B = \langle n_1 n_2 \rangle - \langle n_1 \rangle \langle n_2 \rangle$. Let $W = n_1 n_2$, $Q_1 = n_1$, $Q_2 = n_2$.
Expanding around the means:
\begin{align}
    B &= (\overline{W} + \delta \langle W \rangle) - (\overline{Q_1} + \delta \langle Q_1 \rangle)(\overline{Q_2} + \delta \langle Q_2 \rangle) \nonumber \\
    &\approx (\overline{W} - \overline{Q_1}\overline{Q_2}) + \underbrace{\delta \langle W \rangle - \overline{Q_2}\delta \langle Q_1 \rangle - \overline{Q_1}\delta \langle Q_2 \rangle}_{\text{Linear Term}}.
\end{align}
Squaring the linear term (and noting symmetry between $Q_1$ and $Q_2$):
\begin{align}
    \text{Var}(B) &\approx \frac{1}{D} \left[ \text{Var}_T(W) + \overline{Q_2}^2 \text{Var}_T(Q_1) + \overline{Q_1}^2 \text{Var}_T(Q_2) \right. \nonumber \\
    &\quad \left. - 2\overline{Q_2}\text{Cov}_T(W, Q_1) - 2\overline{Q_1}\text{Cov}_T(W, Q_2) + 2\overline{Q_1}\overline{Q_2}\text{Cov}_T(Q_1, Q_2) \right].
\end{align}
For distinct modes $i \neq j$, $\text{Cov}_T(n_i, n_j) \approx 0$ relative to variances. We retain only self-correlations.
\begin{align}
    \text{Var}(B) &\approx \frac{1}{D} \left[ \text{Var}_T(n_1 n_2) + 2\lambda^2 \text{Var}_T(n_1) - 4\lambda \text{Cov}_T(n_1 n_2, n_1) \right] \nonumber \\
    &= \frac{1}{D} \left[ 3\lambda^4 + 2\lambda^2(\lambda^2) - 4\lambda(\lambda^3) \right] \nonumber \\
    &= \frac{1}{D} \left[ 3\lambda^4 + 2\lambda^4 - 4\lambda^4 \right] \nonumber \\
    &= \frac{\lambda^4}{D}.
\end{align}
\begin{equation}
    \text{Var}(B) \approx \frac{1}{D} \left( \frac{N}{L} \right)^4
\end{equation}

\section{Maximal effective Fisher information}
We will derive Eq.~(20) in this section. From $F_Q\succeq F $ (i.e., $F_Q-F\succeq 0$) it follows that 
\begin{equation}
    \bm{I}^{\text{prior}}+F_Q \succeq \bm{I}^{\text{prior}}+F
\end{equation}
Both $(\bm{I}^{\text{prior}}+F)$ and $(\bm{I}^{\text{prior}}+F_Q)$ are invertible, because $\bm{I}^{\text{prior}}\succ 0$. Therefore, it follows that 
\begin{equation}
    (\bm{I}^{\text{prior}}+F)^{-1} \succeq (\bm{I}^{\text{prior}}+F_Q)^{-1}
\end{equation}.

Eq.~(20) of the main text follows. 

\section{Estimating $F_{\text{eff}}$}
We derive Eqs.~(25), (26) of the main text in this section. If $\lambda_1, \lambda_2, \cdots$ are the eigenvalues of $\bm{I}^{\text{prior}}+F_Q$, we can use the general Weyl inequalities to estimate them in terms of the eigenvalues $F_1, F_2, \cdots, F_{\ell+1}$ of $F_Q$ and $\sigma_1^2, \cdots, \sigma_{\ell}^2$. Without loss of generality, we assume that $F_1\leq F_2\leq ...\leq F_{\ell+1}$ and $\sigma_{\varphi}\leq \sigma_1\leq\cdots\leq \sigma_{\ell}$. Also, we assume $\lambda_1\leq \lambda_2 \leq \cdots, \leq \lambda_{\ell+1}$ The Weyl inequalities read~\cite{bhatia2001linear, thompson1972inequalities, horn2012matrix}:
\begin{equation}
    F_k+\frac{1}{\sigma_{\max}^2}\leq \lambda_k \leq F_{k}+\frac{1}{\sigma_{\min}^2}
\end{equation}
In particular, for extreme values of $k$, 
\begin{equation}
    \begin{split}
        F_{\min}+\frac{1}{\sigma_{\max}^2} &\leq \lambda_{\min} \leq F_{\min}+\frac{1}{\sigma_{\min}^2}\\
        F_{\max}+\frac{1}{\sigma_{\max}^2}&\leq \lambda_{\max} \leq F_{\max}+\frac{1}{\sigma_{\min}^2}\\
    \end{split}
\end{equation}
Eq.~(25) of the main text follows from the above.


\end{document}